\documentstyle[multicol,aps,psfig]{revtex}
\begin{document}

\newcommand{\cd}{\makebox[0.08cm]{$\cdot$}}
\newcommand{\lsim}
{\ \raisebox{2.3pt}{$<$}\hspace{-7.0pt}\raisebox{-2.75pt}{$\sim$}\ }
\newcommand{\gsim}
{\ \raisebox{2.3pt}{$>$}\hspace{-7.0pt}\raisebox{-2.75pt}{$\sim$}\ }

\title{Interplay of soft and hard processes and hadron $p_T$ 
spectra in $pA$ and $AA$ collisions}

\author{Enke Wang$^a$ and Xin-Nian Wang$^{a,b}$}

\address{$^a$ Institute of Particle Physics,
        Huazhong Normal University, Wuhan, 430079, China}
\address{$^b$ Nuclear Science Division, Mailstop 70-319,
        Lawrence Berkeley National Laboratory, Berkeley, CA 94720}

\date{April 6, 2001}

\preprint{LBNL-47702}

\maketitle

\begin{abstract}
Motivated by a schematic model of multiple parton scattering within 
the Glauber formalism, the transverse momentum spectra in $pA$ and $AA$
collisions are analyzed in terms of a nuclear modification factor
with respect to $pp$ collisions. The existing data at the CERN Super 
Proton Synchrotron (SPS) energies are shown to be consistent
with the picture of Glauber multiple scattering in which the interplay 
between soft and hard processes and the effect of absorptive processes
lead to nontrivial nuclear modification of the particle spectra.
Relative to the additive model of incoherent hard scattering, the 
spectra are enhanced at large $p_T$ (hard) by multiple scattering while 
suppressed at low $p_T$ (soft) by absorptive correction 
with the transition occurring at around 
a scale $p_0\sim 1-2$ GeV/$c$ that separates soft and hard processes.
Around the same scale, the $p_T$ spectra in $pp$ collisions also
change from an exponential form at low $p_T$ to a power-law behavior
at high $p_T$. At very large $p_T\gg p_0$, the nuclear enhancement 
is shown to decrease like $1/p_T^2$. Implications of these nuclear 
effects on the study of jet quenching, parton thermalization and 
collective radial flow in high-energy $AA$ collisions are discussed.
\end{abstract}

\begin{multicols}{2}

\section{Introduction}

Hadron yields, spectra and correlations have been the focus of many 
experiments in relativistic heavy-ion collisions. They
provide a snapshot of the state of matter when
hadrons stopped interacting with each other, a stage often
referred to as freeze-out in heavy-ion collisions. One can then 
infer the condition in the early stage prior to the freeze-out.
Such a procedure relies crucially on our understanding of the dynamical
evolution of the system. On the other hand, knowledge of the initial
condition at the beginning of thermalization will also help one 
to unravel the history of evolution. This is especially the case
when a complete thermalization cannot be achieved in certain regions
of phase space (large $p_T$, for example). Therefore, it is important 
to study the hadron spectra in $p+p$ and $p+A$ collisions that will 
help us to understand the initial 
condition of the dense matter in high-energy heavy-ion collisions.

Hadron production at large transverse momentum in $p+p$ collisions 
is relatively well understood within the perturbative QCD (pQCD) parton
model. Because of the large transverse momentum scale, the hard
parton-parton scattering processes can be calculated in pQCD, while the
non-perturbative effects can be factorized into universal parton distributions
and parton fragmentation functions. These parton distributions and 
fragmentation functions can be independently measured
in deep-inelastic electron-nucleon scattering, $e^+e^-$ annihilation
and other processes. The calculated inclusive hadron spectra either in
leading order \cite{owens} or next-to-leading order \cite{bk95} agree
very well with experimental data. 

In $pA$ collisions, one should take into account multiple scatterings inside
the nucleus. Because of the interference effect, 
these multiple scatterings will
give rise to different $A$-scaling behavior of the spectra at different 
values of $p_T$. Generalizing to $AA$ collisions, one should expect
similar behavior for the spectra of initially produced partons. These 
produced partons would undergo further interactions, possibly leading to 
thermalization (or partial thermalization) and expansion, which would be 
reflected in the final hadron spectra. Therefore, one should understand 
first the nuclear modification of the hadron spectra from $p+p$ to $p+A$ 
and to $A+A$ due to initial multiple scattering. Only then can one reliably 
extract information about the freeze-out conditions, e.g., temperature and 
flow velocity \cite{stachel,na49frz}, 
and parton energy loss \cite{loss} from the 
final hadron spectra.

In this paper, we will analyze the nuclear modification of the hadron $p_T$ 
spectra in $p+A$ collisions motivated by a schematic model of multiple 
parton scattering in which coherence and absorptive corrections are 
included via the Glauber multiple scattering formalism.
Multiple scatterings generally enhance the large $p_T$ spectra relative 
to the additive model of hard scattering. It can be shown \cite{xnw-rep}
that absorptive corrections and the power-law behavior of perturbative 
parton cross section are the main reasons why the nuclear 
enhancement at high $p_T$ decreases like $1/p_T^2$. We will also show 
that the same absorptive processes suppress the spectra relative 
to the additive model at low $p_T$, where soft processes dominate and 
the $p_T$ spectra deviate from a power-law behavior. By analyzing the
experimental data in terms of the proposed nuclear modification factor, 
we can phenomenologically determine the $p_T$
scale that separates soft and hard processes by the transition 
from nuclear suppression to nuclear enhancement of the $p_T$ spectra.

\section{Multiple Parton Scattering in $p+A$ Collisions}

Multiple hard parton scatterings in QCD are considered as high-twist
processes and their contributions to the cross section of 
hadronic collisions are generally suppressed by $1/Q^2$,
where $Q$ is the momentum scale of the processes. The
coefficients of the these contributions are generally related to the
multiple parton correlation functions inside a hadron \cite{qiu1}. In
collisions involving a nucleus, such high-twist contributions are
enhanced by $A^{1/3}$ due to the large nuclear size. For certain
quantities like the transverse momentum imbalance of dijets in
photon-nucleus collisions \cite{luo}, parton energy loss in
deep-inelastic lepton-nucleus scattering \cite{xgxw},
broadening of the jet transverse
momentum in deep-inelastic lepton-nucleus scattering \cite{guo1} and 
Drell-Yan pairs in $p+A$ collisions \cite{guo2}, these high-twist
terms are the leading contributions. However, such QCD treatment of
multiple parton collisions is so far limited to the parton level and
has not yet been extended to hadron spectra. In this paper, we will
use instead a schematic Glauber model \cite{petersson}
of multiple parton scattering to motivate our analysis.

Let us denote $h^{j}_{iN}$ as the differential cross
section for a parton-nucleon scattering, $i+N \to j+X$,
\begin{equation}
h^{j}_{iN}=\sum_{b,c} \int dx_b f_{b/N}(x_b)
\frac {\hat s}{\pi} \frac {d\hat
{\sigma}_{ib\rightarrow jc}}{d\hat t} \delta (\hat {s} +\hat {t} +\hat {u})
\end{equation}
where $f_{b/N}(x_b)$ is the parton distribution function inside
a target nucleon, $\hat s$, $\hat t$ and $\hat u$ are Mandelstam variables
and $d\hat {\sigma}_{ib\rightarrow jc}/d\hat t$ are the differential
parton-parton cross sections. In terms of $h^{j}_{iN}$, we define the
effective parton-nucleon total cross section as
\begin{equation}
\sigma_{iN}(p_i)=\frac{1}{2} \sum_j \int \frac {d^3p_j}{E_j}
h^{j}_{iN}(p_i,p_j),
\end{equation}
and the differential nucleon-nucleon cross section for parton production as
\begin{equation}
E\frac {d\sigma_{NN}^j}{d^3p}
=\sum_i \int \frac {dx_i}{x_i} f_{i/N}(x_i) h^j_{iN}(p_i,p) .
\end{equation}
The effective parton-nucleon and nucleon-nucleon cross sections are 
only finite after one introduces some effective infrared cut-off in the
parton-parton scattering processes. We will discuss this cut-off later. 
Following the approach by \cite{petersson,xnw-rep}, in which the Glauber 
approximation is used for multiple parton scattering, one can find
the parton spectra in $pA$ collisions up to double scattering 
approximation as
\begin{eqnarray}
        E\frac{d\sigma^j_{NA}}{d^3pd^2b} &\approx& 
        E\frac {d^3\sigma_{NN}^j}{d^3p} T_A(b) +
        \frac{1}{2} T^2_A(b) \sum_i \int 
        \frac {dx_i}{x_i} f_{i/N}(x_i) \nonumber \\ 
       &\times&\left[ \sum_k\int\frac{d^3p_k}{E_k}
        h^k_{iN}h^j_{kN} - (\sigma_{iN}+\sigma_{jN})h^j_{iN}\right],
\end{eqnarray}
where $T_A(b)$ is the nuclear thickness function with normalization
$\int d^2bT_A(b)=A$. We take the nucleus to be a hard sphere of radius
$r_A=r_0A^{1/3}$ with $r_0=1.14$ fm.
One can then obtain the ratio of the differential cross sections
of $N+A\rightarrow j+X$ and $N+N\rightarrow j+X$,
\vspace{0.5in}
\begin{eqnarray}
R_A &\equiv& \frac{Ed^3\sigma^j_{NA}/d^3p}{AEd^3\sigma^j_{NN}/d^3p}
      \nonumber \\
      &=& 1 + \frac {9A^{1/3}}{16\pi r^2_0 Ed\sigma^j_{NN}/d^3p}
      \sum_i \int \frac {dx_i}{x_i} f_{i/N}(x_i)
       \nonumber \\
     &\times &\left[ \sum_k \int \frac {d^3p_k}{E_k} h^k_{iN}h^j_{kN}
     - (\sigma_{iN}+\sigma_{jN}) h^j_{iN} \right].
\end{eqnarray}
Compared to the additive model of parton scattering which gives
$R_A=1$, the second term in the above equation gives us the nuclear
modification of the parton spectrum due to multiple parton scattering
inside a nucleus. This term contains contributions from both the
double parton scattering and the absorptive correction to the single
scattering processes. Notice that the absorptive contribution at this
order is negative. As we will demonstrate, it is the cancellation by
the absorptive correction that leads to many interesting and
nontrivial features of the nuclear modification of the particle
spectra.

\section{Nuclear Modification: A Schematic Study}

As a demonstration of the consequences of multiple parton scattering,
absorptive correction and the resultant interplay between low 
and high $p_T$ behavior of the particle spectra in $pA$ collisions, 
we illustrate the nuclear modification of parton spectra in a schematic
model. Assuming parton-hadron duality, the conclusions can be applied 
qualitatively to hadron spectra.

For a schematic study, let us assume that all partons are identical and
the differential parton-nucleon cross sections have a simple
regularized power-law form in $p_T$,
\begin{equation}
h^j_{iN}\equiv h(p_T)=\frac{C}{(p_T^2+p_0^2)^n}, \;\;(|y|<\Delta Y/2),
\end{equation}
with the total parton-nucleon cross section,
\begin{equation}
\sigma_{iN}\equiv \sigma =\int dyd^2p_T
\frac{C}{(p_T^2+p_0^2)^n}=\frac{\pi\Delta Y C}{(n-1)p_0^{2n-2}}.
\end{equation}
Such a form is motivated by the fact that both the pQCD calculation
and experimentally 
measured differential jet production cross sections show such a 
power-law behavior. The parameter $p_0$ is introduced as an
infrared cut-off. It can be regarded phenomenologically as a scale that
separates soft and hard processes, because for $p_T<p_0$ the parton
spectra deviate significantly from a power-law behavior.

\begin{figure}
\centerline{\psfig{figure=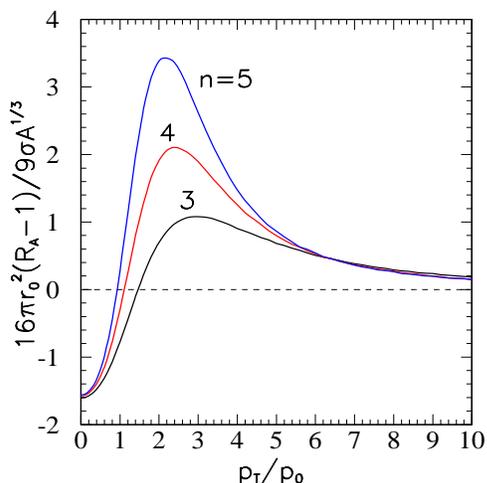,width=2.5 in,height=2.5in}}
\caption{The nuclear modification factor as a function of $p_T$ for
different values of $n$ in a schematic model of multiple parton
scattering with a simple form of parton-nucleon cross section 
$1/(p_T^2+p_0^2)^n$.}
\label{fig:schem}
\end{figure}

Using this schematic parton-nucleon cross section, the 
nuclear modification factor can be simplified as
\begin{eqnarray}
R_A&=&1+\frac{9}{16}A^{1/3}\frac{\sigma}{\pi r_0^2} \left\{
\frac{(n-1)}{\pi}\left( 1 + \frac{p_T^2}{p_0^2} \right)^n \right.
\nonumber \\
&\times&\left. \int d^2y_T (1+y_T^2)^{-n} 
[1+(\vec{p}_T/p_0-\vec{y}_T)^2]^{-n} -2 \right\}.
\end{eqnarray}
This modification factor as a function of $p_T/p_0$ for different values 
of $n$ is shown in Fig.~\ref{fig:schem}, where we plot $R_A-1$ in 
units of the modification strength $9A^{1/3}\sigma/(16\pi r_0^2)$.
One can also evaluate the nuclear modification factor analytically 
at two different limits of $p_T$,
\begin{eqnarray}
R_A=1+\frac{9}{16}A^{1/3}\frac{\sigma}{\pi r_0^2} 
& & \left[ \frac{2n^2}{n-2} \frac{p_0^2}{p_T^2}
+ {\cal O} (\frac{p_0^4}{p_T^4})\right], \nonumber \\
 & &(p_T \gg p_0) ; \nonumber \\
R_A=1+\frac{9}{16}A^{1/3}\frac{\sigma}{\pi r_0^2}
& & \left[-\frac{3n-1}{2n-1} \right. \nonumber \\
& & +\frac{n(n-1)(2n+3)}{2(4n^2-1)} \frac{p_T^2}{p_0^2} \nonumber \\
& & \left. + {\cal O} (\frac{p_T^4}{p_0^4})\right],(p_T \ll p_0).
\label{eq:rlimit}
\end{eqnarray}

As one can see from both Fig.~\ref{fig:schem} 
and the above asymptotic behavior, the parton
spectra are enhanced at large $p_T$ due to contribution from double scattering.
However, due to the cancellation by the absorptive correction to the single
scattering, the enhancement decreases like $p_0^2/p_T^2$ at large $p_T$. This
asymptotic behavior will never occur if one does not include the absorptive
corrections which are embedded in the Glauber formalism employed in this model.
At low $p_T$, contributions from double scattering become smaller than the
absorptive correction, so that the effective parton spectra are reduced
($R_A<1$) relative to the additive model. For very large values of the $pp$ 
cross section and the size of the nucleus $A$, we notice that the modification
factor $R_A$ could become negative in Eq.~(\ref{eq:rlimit}) at $p_T=0$,
signaling the breakdown of the double scattering approximation. In this case,
one would have to include triple and other multiple scatterings. 
One can show along
the same line as the double scattering, the contribution from the sum of 
triple scattering and the corresponding absorptive correction is positive
at $p_T=0$ and is proportional to $(A^{1/3}\sigma)^2$. Assuming one can
exponentiate the sum of all multiple scatterings and absorptive corrections,
the modification factor at $p_T=0$ would take the form,
\begin{equation}
R_A(p_T=0)\approx \frac{1}{A\sigma}\int d^2b 
\left[ 1-e^{-c\sigma T_A(b)} \right] \; . \label{eq:rpt0}
\end{equation}
In this regime, the absorptive parts dominate and $R_A \propto 1/A^{1/3}$.
This is consistent with the wounded nucleon model for soft particle 
production. The coefficient $c\leq 1$ is a result of partial cancellation 
by contributions from multiple scattering. 

It is clear that the shape of the nuclear modification of the inclusive 
spectra results from the interplay between multiple scattering and absorptive
corrections. This is especially true at low $p_T$. Even though it gives the
apparent broadening of the effective spectra in $pA$ relative to $pp$
collisions, the underlying physics is very different from the random-walk 
model \cite{satz} for the $p_T$ broadening of soft hadrons. One can
expect the same behavior in $A+A$ collisions if there is no additional nuclear
effect, such as final state rescattering or jet quenching due to
radiative parton energy loss. Analyzing the experimental data in this 
framework  with respect to the above baseline behavior would allow 
one to identify and study new effects caused by the presence of dense matter.

Another important feature of the modification factor is its sensitivity
to the form of particle spectra in $pp$ collisions. At low $p_T$,
its behavior is not very sensitive to the form of differential
parton-nucleon cross section. If a Gaussian form of the spectrum is usded, the
modification factor remains roughly the same at $p_T\sim 0$. In contrast,
the high-$p_T$ behavior of the enhancement, $\sim 1/p_T^2$, is strictly the
consequence of the power-law form of the parton-nucleon cross section. In 
fact, with a Gaussian form, the enhancement will increase with $p_T^2$
exponentially. Therefore, the observed large $p_T$ behavior of the Cronin 
enhancement \cite{cronin} is consistent with the power-law form of the jet
production cross section and with the picture of multiple parton scattering in
$p+A$ collisions. The nuclear modification factor at large $p_T$ also 
depends on the power $n$ of the $pp$ spectra. Since the power $n$ decreases
with energy, one should expect $R_A$ to decrease with energy. This trend
has been observed in experiments \cite{straub} in the energy range
$\sqrt{s}=$20--40 GeV.

Overall the nuclear modification factor $R_A$ has nontrivial and 
interesting $p_T$ dependence. It is smaller than 1 at low $p_T$ and 
larger than 1 at intermediate $p_T$. The transition ($R_A=1$) occurs 
at around $p_T\sim p_0/\sqrt{n}\sim \sqrt{\langle p_T^2\rangle}$.
It approaches 1 again at very large $p_T\gg p_0$. The modification 
strength at both low and high $p_T$ is proportional to the size of 
the nucleus. As we will see in the next section, this is qualitatively 
consistent  with the experimental data of inclusive hadron spectra 
in $pA$ collisions. Since $p_0$ is a momentum scale that set the 
onset of power-law-like spectra in $pp$ collisions, one can consider 
it a scale separating soft and hard processes. According to the 
schematic model, one should be able to determine independently
this scale from the nuclear modification factor of the hadron spectra.

\section{Analysis of experimental data}

\begin{figure}
\centerline{\psfig{figure=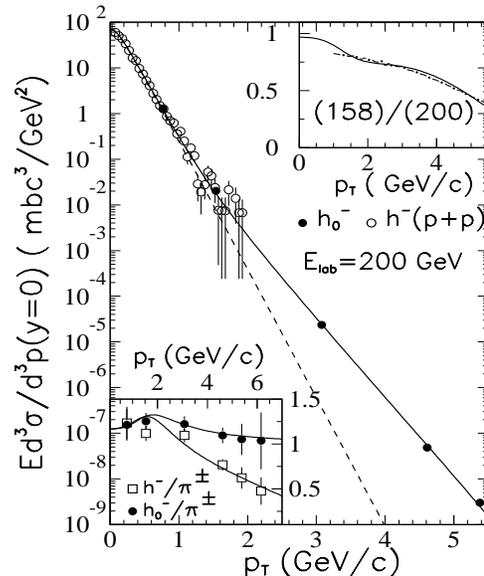,width=2.5 in,height=3.0in}}
\caption{Negative hadron spectra in $pp$ collisions 
at $E_{\rm lab}=200$ GeV from Refs.\protect\cite{cronin-ex1,na35} and the
fit (solid line) according to Eq.~(\protect\ref{eq:para}) with fit parameters
given in Table ~\protect\ref{table}. The dashed line is the underlying
exponential component. The $h_0^-$ is defined as
$h_0^-=(\pi^+ +\pi^-)/2+K^-+\bar{p}$. The upper inserted box shows 
the ratio of parameterizations (solid line) at $E_{\rm lab}=158$ 
and 200 GeV and the corresponding pQCD parton model 
calculation (dot-dashed line). The lower box shows the ratio
of $h^-/\pi^\pm$ and $h_0^-/\pi^\pm$.}
\label{fig:parapp}
\end{figure}

To facilitate the analysis of $pA$ and $AA$ spectra in terms of the nuclear
modification factor $R_A$, one needs to know the $pp$ spectra at the same
energy. Since we will have to take ratios of hadron spectra from different
experiments with different $p_T$ bins, this can only be achieved by
parameterizing the baseline spectra. Shown in Fig.~\ref{fig:parapp} are
the measured negative hadron spectra \cite{cronin-ex1,na35} and the
parameterization in $pp$ collisions at $E_{\rm lab}=200$ GeV. We use
a two-component parameterization,
\begin{equation}
f(p_T)=C_0 e^{-m_T/T_0} + C \frac{(1-2p_T/\sqrt{s})^a}{(p_T^2+p_0^2)^n} .
\label{eq:para}
\end{equation}
The fit parameters are listed in Table~\ref{table} for different
spectra at different energies. The parameters in the
exponential form are mainly determined by the spectra below $p_T<1$ GeV/$c$
while those in the power-law form are mainly determined by the large
$p_T$ region. Shown as a dashed line is the contribution from the
exponential component. It is clear that at $p_T>2 $ GeV/$c$, the spectrum
is already dominated by the power-law behavior. The power-law form we use for
the CERN-SPS energy range contains a factor $(1-2p_T/\sqrt{s})^a$ that 
is caused by the rapid decrease of quark distributions at large $x\sim 1$. 
This factor will become negligible at higher energies 
when $2p_T/\sqrt{s}\ll 1$.

\end{multicols}
\begin{table}
\caption{Fit parameters of the hadron spectra in pp collisions}
\label{table} 
\begin{tabular}{|c|c|c|c|c|c|c|}
  & $C_0$ (mbGeV$^{-2}$) & $T_0$ (MeV) & $C$ (mbGeV$^{-2}$) 
& $a$ & $p_0$ (GeV/$c$) & $n$ \\ \hline
$h^- (pp, E_{\rm lab}=158$ GeV) & 169.6 & 154 
& 24565 & 9.3 & 2.43 & 6.29 \\ \hline
$h_0^- (pp, E_{\rm lab}=200$ GeV) & 174.7 & 154 
& 13653 & 9.1 & 2.27 & 6.22 \\ \hline
$h^- (pp, E_{\rm lab}=200$ GeV) & 174.7 & 154 
& 17653 & 9.8 & 2.27 & 6.29 \\ \hline
$\pi^{\pm} (pp, E_{\rm lab}=200$ GeV) & 150.3 & 154 
& 13653 & 9.1 & 2.37 & 6.22 \\ \hline
$h^{\pm} (p\bar{p}, \sqrt{s}=200$ GeV) & 440 & 154 
& 653 & 0 & 1.75 & 4.98 \\
\end{tabular}
\end{table}
\begin{multicols}{2}

In the original UA1 parameterization \cite{ua1}, a power-law form
$A_0/(p_T+p_0)^n$ is used to fit the hadron spectra over a large
range of energies. While this single power-law is sufficient to fit
the spectra for $p_T>0.2$ GeV/$c$, an exponential term is necessary 
to fit the spectra at low transverse momentum $p_T<0.2$ GeV/$c$.

In order to compare to the spectra in $pA$ and $AA$ collisions for
near isospin symmetrical nuclei, the data shown in Fig.~\ref{fig:parapp}
at large $p_T$ are for ''negative'' hadrons with the $\pi^-$ contribution
being replaced by an isospin averaged value,
$h_0^-=(\pi^+ +\pi^-)/2+K^- +\bar{p}$.
Since particle production at large $p_T$ is dominated by the
leading hadrons from valence quark scattering and there are more up-quarks
than down-quarks in a $pp$ system, one should see more $\pi^+$ 
than $\pi^-$ at large $p_T$. Shown in the inserted box at the lower-left
corner are the ratios of $h^-/\pi^{\pm}$ and $h_0^-/\pi^{\pm}$ 
[$\pi^{\pm}=(\pi^+ + \pi^-)/2$]. Note that $h^-/\pi^{\pm}$ decreases with $p_T$
as expected while $h_0^-/\pi^{\pm}$ remains relatively constant. 
At low $p_T$, $h^-$ and $h_0^-$ are approximately the same within
a few percent accuracy. This isospin dependence can be described 
well by the pQCD parton model \cite{xnw00}.
With $h_0^-$ and $h^-$, one will be able to estimate the isospin effect
in $pA$ and $AA$ collisions. In the inserted box at the upper-right
corner we also plot the ratio of the hadron spectra in $pp$ collisions at
$E_{\rm lab}=158$ and 200 GeV. Since there are no experimental data for
$pp$ collisions at $E_{\rm lab}=158$ GeV, we parameterize the spectrum
at $E_{\rm lab}=158$ GeV to fit the ratio (dot-dashed line) obtained 
from a pQCD parton model calculation \cite{xnw00}.

\begin{figure}
\centerline{\psfig{figure=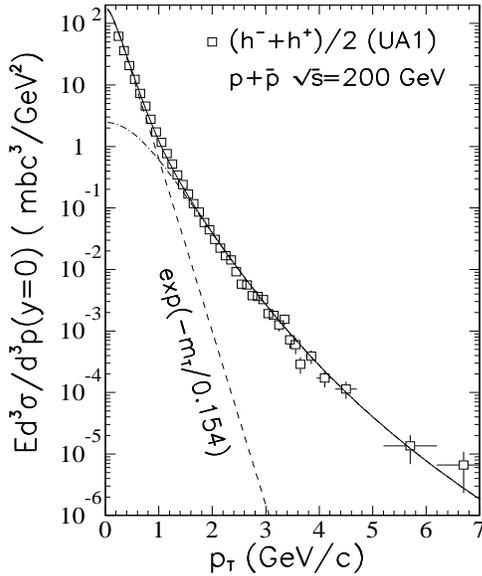,width=2.5 in,height=3.0in}}
\caption{Charged hadron spectra in $p\bar p$ collisions 
at $\sqrt{s}=200$ GeV from Ref.\protect\cite{ua1} and the
fit (solid line) according to Eq.~(\protect\ref{eq:para}) 
with fit parameters given in Table ~\protect\ref{table}. 
The dashed line is the underlying exponential component
and the dot-dashed line is the power-law component.}
\label{fig:parappbar}
\end{figure}

At the BNL Relativistic Heavy-ion Collider (RHIC) energy, we expect 
the power-law contribution to become more
important. Shown in Fig.~\ref{fig:parappbar} is the charged hadron
spectrum in $p\bar{p}$ collisions at $\sqrt{s}=200$ GeV \cite{ua1}
and the parameterization. The power-law is indeed more prominent
than at the CERN-SPS energy. It becomes dominant already at 
around $p_T=1.5$ GeV/$c$. There is a general trend that both $p_0$ 
and $n$ of the power-law component decrease with colliding energy.
However, the underlying exponential term remains the same. 

With these parameterizations of hadron spectra in $pp$ collisions, we
can analyze the hadron spectra in $pA$ and $AA$ collisions in terms of
a nuclear modification factor which is defined in general as
\begin{equation}
R_{AB}(p_T)=\frac{d\sigma_{AB}/dyd^2p_T}
{\langle N_{\rm binary}\rangle d\sigma_{NN}/dyd^2p_T}
\end{equation} 
for $AB$ collisions, where 
$\langle N_{\rm binary}\rangle=\int d^2b T_{AB}(b)$ is the
number of binary collisions averaged over the impact-parameter
range of the corresponding centrality. Here $T_{AB}(b)$ is the
nuclear overlap function for $AB$ collisions at impact parameter $b$.
For minimum-biased events,
$\langle N_{\rm binary}\rangle=AB$. For the purpose of the study
in this paper, we will select event centrality according to the
geometrical cross sections.

\begin{figure}
\centerline{\psfig{figure=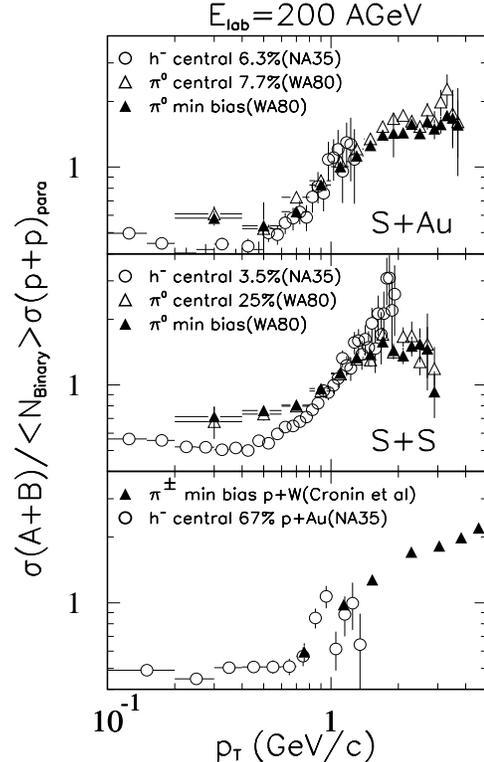,width=2.5 in,height=4.0in}}
\caption{The nuclear modification factor $R_{AB}(p_T)$ for
hadron spectra in $pA$ and $AB$ collisions at $E_{\rm lab}=200$ GeV. 
The data are from Refs.\protect\cite{cronin-ex1,na35,wa80}.}
\label{fig:ratio1}
\end{figure}

\begin{figure}
\centerline{\psfig{figure=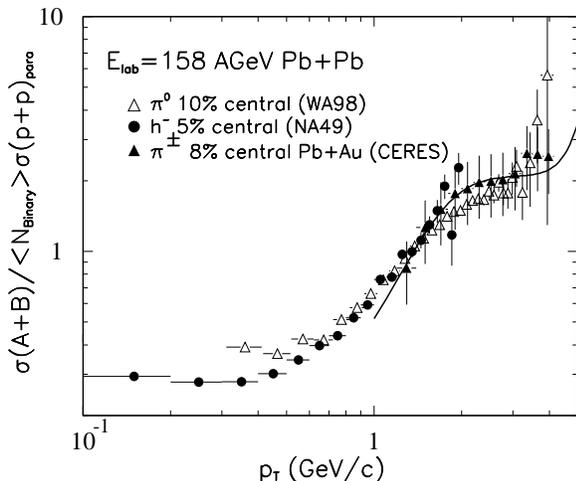,width=3.0 in,height=2.5in}}
\caption{The nuclear modification factor $R_{AB}(p_T)$ for hadrons
in $AB$ collisions at $E_{\rm lab}=158$ GeV. The data are from 
Refs.\protect\cite{wa98,na49,ceres}. The line is the pQCD parton model
calculaiton.}
\label{fig:ratio2}
\end{figure}

Shown in Figs.~\ref{fig:ratio1} and \ref{fig:ratio2} are the nuclear 
modification factors for hadron spectra in $pA$ and $AB$ collisions with
different centralities at CERN-SPS energies. One can see that 
both $pA$ and $AB$ data have features very similar to those shown by
the schematic model. The nuclear modification factors are smaller than 1
at low $p_T$. According to Eq.~(\ref{eq:rpt0}), the modification factor
at $p_T=0$, $R_{AB}(0)$, should decrease with increasing nuclear size.
This is indeed seen in the data. As $p_T$ increases, $R_{AB}$ also 
increases and becomes larger than 1 at
about $p_T=1-2$ GeV/$c$. According to the physical picture depicted
by the schematic model, this momentum scale represents the onset of
hard processes underlying the hadron spectra. The increase of $R_{AB}$
is then due to the onset of the hard component of incoherent parton
scattering underlying the hadron production mechanism as $p_T$ increases.
The data then indicate that at $p_T=1-2$ GeV/$c$, the hard component
becomes completely dominant. Indeed, such a value
is consistent with the momentum scale around which the hadron spectra
in $pp$ collisions begin to be dominated by a power-law behavior of hard
parton scatterings. In principle, the nuclear modification factor
should peak at intermediate values of $p_T$ and decrease again to 
approach to 1 at very large $p_T$. However, at $p_T=4$ GeV/$c$,
one is already close to the kinetic limit of $\sqrt{s}/2$ at
the CERN-SPS energy. Near the kinetic limit the spectra become
sensitive to other nuclear effects such as the Fermi motion that will
increase the parton distribution function at $x\sim 1$, leading
to increase of $R_{AB}$ again. In $pA$ collisions at the RHIC energy, 
one will certainly see the decrease of $R_A$ again at 
large $p_T \gg 2$ GeV/$c$.

The increase of the nuclear modification factor with $p_T$ due to
the onset of hard parton scatterings appears very similar to the
broadening of hadron spectra due to collective radial flow in a 
hydrodynamic picture. If there are strong interactions among
partons in the early stage and hadrons in the late stage of heavy-ion
collisions, the system will be driven to local thermal equilibrium.
The information of initial multiple parton scattering contained
in the initial parton spectra could be partially or completely
erased by the thermalization. Collective flow then will be developed 
and hadron spectra will become broader than in $pp$ and $pA$ collisions 
due to the boost by a collective flow velocity \cite{hydro}. Such an
effect essentially will also increase $R_{AB}$ as a function of $p_T$. 
The question is how one can distinguish these two apparently 
different dynamics that produce the same final hadron spectra.   
Mass dependence of the hadron spectra has been proposed as a
unique measure of the collective flow effect in heavy-ion 
collisions \cite{nxu}. Since a collective flow provides a common
velocity boost for all particles, heavy particles will then acquire
more transverse momentum in the nonrelativistic region 
($p_T$\lsim$ m_h$). One then should see a linear
mass dependence of the slope parameter from an exponential fit
of the measured hadron spectra \cite{nxu}. In the current parton 
model\cite{xnw00}, one cannot exclude the mass dependence of 
the nuclear modification factor. It is therefore important to
perform the current analysis for different hadron species in
$pp$ and $pA$ collisions to find out whether nuclear modification
factors have a mass dependence. Only then can one find out whether
there is complete or partial thermalization and to what extent
the effect of initial multiple parton scatterings has survived
the final state interaction and contributed to the apparent
collective radial flow.

\section{High $p_T$ spectra and parton energy loss}

Since the interaction for an energetic parton inside a medium is 
dominated by small angle forward scatterings, thermalization will
be less complete for high $p_T$ partons. Hydrodynamic description
will then become less relevant. Large $p_T$ hadron spectra will be 
determined by how an energetic parton propagates inside the medium.
Since the large $p_T$ hadron spectra are calculable in the pQCD parton 
model, they are good probes of the parton dynamics in dense matter.
The nuclear modification factor at large $p_T$ is a convenient
and efficient way to determine the effects of the final state
interaction of energetic partons inside a medium.

\begin{figure}
\centerline{\psfig{figure=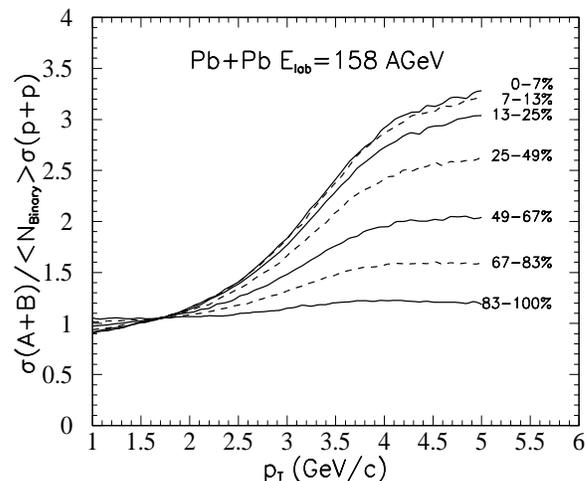,width=3.0 in,height=2.5in}}
\caption{The predicted nuclear modification factor $R_{AB}(p_T)$ for
$\pi_0$ spectra in $Pb+Pb$ collisions at $E_{\rm lab}=158$ GeV with
different centrality cuts.}
\label{fig:ratio3}
\end{figure}

To demonstrate whether the experimental data at the CERN-SPS
are consistent with the picture of multiple parton scattering,
we also plot in Fig.~\ref{fig:ratio2} the pQCD-inspired parton model
calculation of the nuclear modification factor \cite{xnw00}. 
In such a model, one extends the collinear factorized parton model 
to include intrinsic transverse momentum and its broadening due 
to multiple scattering in nuclear matter. The value of the
intrinsic transverse momentum and its nuclear broadening are adjusted once
and the model can reproduce most of the experimental data in $pp$ and $pA$ 
collisions \cite{xnw00}. In $AA$ collisions, the effect of parton
energy loss is modeled by the modification of parton fragmentation 
functions \cite{whs} in which the distribution of leading hadrons 
from parton fragmentation is suppressed due to parton energy loss. 
We will not describe the model here and refer readers to Ref.~\cite{xnw00}
for detail. We note that the behavior of the nuclear modification 
factor at $p_T>2.0$ GeV/$c$ in $Pb+Pb$ collisions is well described 
by the model without any additional final state medium effect such 
as the parton energy loss or jet quenching \cite{loss}. Since the
pQCD parton model cannot deal with soft processes, the calculations
are only reliable for $p_T>2$ GeV/$c$. The $p_T$ broadening due to 
multiple parton scattering is essential to account for the $p_T$ 
dependence of the nuclear modification factor. Different variations
of the parton model studies \cite{mgpl} give the same conclusion.
Such a good agreement 
is not just a coincidence. The calculated absolute differential cross
section for different colliding systems ($S+S$, $S+Au$ and $Pb+Pb$)
and for different centralities also agree well with the
experimental data \cite{xnw00}. Shown in Fig.~\ref{fig:ratio3} are 
the calculated nuclear modification factors for $\pi^0$ spectra 
in $Pb+Pb$ collisions with different centrality cuts, assuming no
parton energy loss. Though the
$p_T$ dependences of the modification factors are similar, the
magnitudes increase with centrality. The impact-parameter ranges for
different centralities are determined by the fractions of geometrical
nuclear cross sections.

Recent theoretical studies of parton propagation in a dense medium
predict a substantial parton energy loss due to induced gluon radiation
as the parton interacts with the dense medium \cite{GW1,BDPS,BGZ,GLV,wied}.
If a dense medium is produced in heavy-ion collisions, large $p_T$
jets will lose energy as they propagate through the medium. This
will lead to suppression of large $p_T$ hadrons \cite{loss}. However,
as shown by Fig.~\ref{fig:ratio2}, the experimental data of
central $Pb+Pb$ collisions at the CERN-SPS are consistent with
the picture of initial multiple parton scattering without any
parton energy loss in the final state. Careful analysis of the data
against the pQCD parton model can actually exclude effects of any significant
amount of parton energy loss in central $Pb+Pb$ collisions at the
CERN-SPS \cite{xnw98a}. To be consistent with the current theoretical
estimates of parton energy loss in a dense medium, these data imply that
the initial energy density would be small and the lifetime of the
dense system is very short in $Pb+Pb$ collisions at the CERN-SPS. On the
other hand, preliminary data from RHIC experiments \cite{drees}
have shown significant suppression of high $p_T$ hadron spectra.
Combined with the null results on hadron suppression in heavy-ion
collisions at the CERN-SPS, RHIC data would imply the onset of parton
energy loss due to the increasing initial energy density and the 
lifetime of the dense system created in heavy-ion collisions
at RHIC energies. Study of the nuclear modification factors for
hadron spectra at large $p_T$ will help to determine the average
parton energy loss in the medium produced in heavy-ion 
collisions \cite{loss}.

The nuclear modification factor will also help us to understand
the energy dependence of the parton energy loss. Earlier theoretical
studies \cite{GW1,BDPS,BGZ} gave a constant or a weakly 
energy-dependent parton energy loss $dE/dx$. However, more recent
studies in an opacity expansion approach \cite{GLV,wied} predict
a stronger energy-dependence for $dE/dx$ when the parton energy
is small ($E<10 \mu$, $\mu$ being the Debye screening mass in the medium).
Parameterization of the numerical results of Ref.~\cite{GLV} gives
$dE/dx \propto E^{1.5}/(10+E)^{1.25}$ for $\mu=0.5$ GeV. The
coefficient depends linearly on the initial parton density and
the system size. This form has a strong energy dependence for
$E\stackrel{<}{\sim} 10$ GeV and a weak dependence at large $E$.

\begin{figure}
\centerline{\psfig{figure=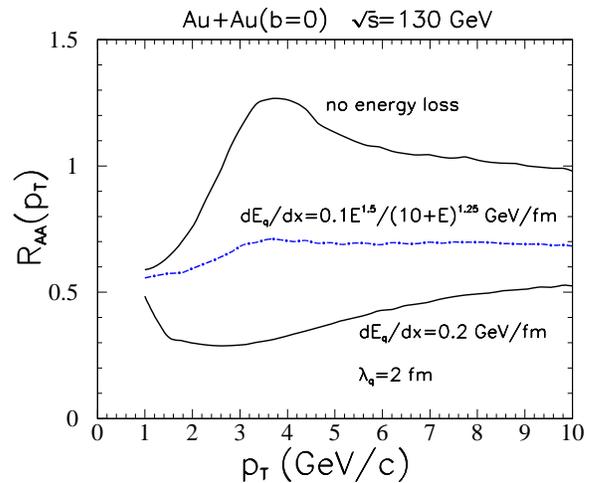,width=3.0 in,height=2.5in}}
\caption{The parton model calculation of nuclear modification factor 
$R_{AB}(p_T)$ for hadrons in central $Au+Au$ collisions at $\sqrt{s}=130$ 
GeV with different forms of parton energy loss. $\lambda_q$ is the value
of mean free path for a propagating quarks used in the calculation.}
\label{fig:ratio4}
\end{figure}
Shown in Fig.~\ref{fig:ratio4} are the nuclear modification factors
calculated in a pQCD parton model \cite{xnw00} for different forms
of parton energy loss. The nuclear modification factor is
suppressed to become smaller than 1 for nonvanishing parton energy
loss. For a constant $dE/dx$, the modification factor increases
with $p_T$ after the initial drop. This is because that the constant
energy loss becomes relatively less important for higher initial parton 
energy. For asymptotically large $p_T$, the finite and constant energy
loss would become negligible and the modification factor $R_{AB}$ would 
approach 1. However, for an energy loss that has a strong energy 
dependence, the shape of the modification factor is very different.
If this is the case, the energy loss for a parton in the 
small $p_T$ region is still
very small and only becomes sizable at large $p_T$. Consequently,
the effect of energy loss is also very small. The modification factor
will then increase, following the trend of the initial parton spectra
due to the transition from soft to hard scatterings. However, when
larger energy loss sets in at higher $p_T$, the hadron spectra are
strongly suppressed leading to much smaller values of $R_{AB}$.
We should note that the $dE/dx$ used in these calculations
is only representative of its value averaged over the entire 
evolution of the dense system. Since the energy loss is
directly proportional to the parton density of the medium
which decreases very fast due to rapid expansion, a small
averaged $dE/dx$ still corresponds to large initial energy loss
and large initial parton density.

Similar to the interplay between soft and hard processes in the 
initial parton production, there should also be a smooth transition 
between effects of hydrodynamic evolution at low $p_T$ and 
parton energy loss at high $p_T$. The hydrodynamic picture is relevant
only if there is local thermalization, which is more likely for
low $p_T$ partons. It breaks down for very large $p_T$ partons because 
the small angle scatterings in QCD are not sufficient to bring 
these partons in equilibrium with the rest of the system. These
high $p_T$ partons will suffer energy loss as they propagate
through the medium leading to suppression of large $p_T$ hadrons.
However, at intermediate $p_T$, one cannot neglect the effect
of partial thermalization or the detailed balance. For example, 
one should consider both gluon absorption and radiation by
a propagating parton. So far all theoretical studies have only 
considered gluon radiation. If gluon absorption is also included,
the effective parton energy loss will be reduced \cite{ewxw}. Since
the effect of gluon absorption will be small for high energy partons 
($p_T\gg T$, $T$ being the temperature of the thermal medium), the
absorption will further increase the energy dependence of the
effective parton energy loss. The $p_T$ dependence of the nuclear
modification factor $R_{AB}(p_T)$ at intermediate $p_T$
will then provide useful information about the thermalization
of the dense medium.

\section{Conclusions}

In this paper we have proposed to analyze the hadron transverse
momentum spectra in terms of the nuclear modification factor
$R_{AB}(p_T)$ which is defined in such a way that a naive additive model
of incoherent hard parton scattering would give $R_{AB}=1$.
We demonstrated in a schematic model of Glauber
multiple parton scattering that the modification factor $R_{AB}(p_T)$
has a nontrivial $p_T$ dependence due to the absorptive processes
and the interplay between soft and hard parton scattering, excluding
final state scatterings. Because of the absorptive processes, the hadron 
production at small $p_T\sim 0$ is coherent and the hadron spectra 
in $AB$ collisions are proportional to the number of participant 
nucleons, leading to $R_{AB}<1$. At large $p_T$ the hard parton
scatterings become incoherent. Multiple parton scatterings then
enhance the hadron spectra so that $R_{AB}>1$. The momentum scale
$p_0$ at which the transition occurs [$R_{AB}(p_0)=1$] can be
identified as the scale that separates soft and hard processes
underlying both $pp$ and $AB$ collisions. Analyses of the existing
experimental data on $pp$, $pA$ and $AB$ collisions indicate 
that $p_0\approx 1-2$ GeV$/c$.

We pointed out that such analyses of future experimental data are
important to study the effect of final state interactions. At low
$p_T$, collective radial flow from hydrodynamic expansion gives
similar $p_T$ dependence of $R_{AB}$. Disentangling the effects of
initial multiple scattering and the radial flow would require
a careful study of the modification factor in $pA$ collisions,
especially its dependence on the hadron mass. At large $p_T$,
parton energy loss will lead to suppression of the hadron spectra.
Experimental measurement of $R_{AB}$ will provide important
information on the initial parton density that is produced in
heavy-ion collisions. At the intermediate $p_T$, we have shown
that the $p_T$ dependence of $R_{AB}$ is sensitive to the
energy dependence of $dE/dx$. This in turn is related to gluon
absorption by the propagating partons reflecting the detailed 
balance in an equilibrating system.

Although experimental data at the CERN-SPS \cite{xnw98a} do not 
indicate any sign of parton energy loss, recent preliminary
experimental data from RHIC \cite{drees} show significant
suppression of large $p_T$ hadrons. Detailed studies of the 
nuclear modification factor $R_{AB}(p_T)$ over the whole range of
$p_T$ in both $pA$ and $AA$ collisions, will help us not only to
determine the parton energy loss or the initial parton density but
also to find out the degree of thermalization and radial
collective flow. Recent studies \cite{v2} also pointed out that 
parton energy loss can also lead to azimuthal anisotropy in 
high $p_T$ hadron spectra which is significantly different
from hydrodynamic behavior. A combination of studies on 
$R_{AB}$ and the azimuthal anisotropy at high $p_T$ will provide
a window to the initial condition and dynamics of early evolution
of the dense matter that was never possible before.

\section*{ACKNOWLEDGMENTS}
We thank M. Gyulassy and I. Vitev for discussions about the energy
dependence of parton energy loss.
This work was supported by  
the Director, Office of Energy 
Research, Office of High Energy and Nuclear Physics, 
Division of Nuclear Physics, and by the Office of Basic Energy Science, 
Division of Nuclear Science, of  the U.S. Department of Energy 
under Contract No. DE-AC03-76SF00098, and by
the National Natural Science Foundation of China under
project 19928511 and 19945001.

\end{multicols}

\end{document}